\documentclass[prb,twocolumn,superscriptaddress,showpacs]{revtex4}
\usepackage[dvips]{graphicx}
\usepackage{amsmath}
\usepackage{amssymb}
\usepackage{epsfig}
\usepackage{wasysym}
\usepackage{subfigure}
\newcommand \be{\begin{equation}}
\newcommand \ee{\end{equation}}
\newcommand \bes{\begin{equation*}} 
\newcommand \ees{\end{equation*}}
\newcommand \bea{\begin{eqnarray}}
\newcommand \eea{\end{eqnarray}}
\newcommand \beas{\begin{eqnarray*}} 
\newcommand \eeas{\end{eqnarray*}}
\newcommand \bfg{\begin{figure}}
\newcommand \efg{\end{figure}}
\newcommand \bfgs{\begin{figure*}} 
\newcommand \efgs{\end{figure*}}
\newcommand \bwt{\begin{widetext}} 
\newcommand \ewt{\end{widetext}}

\newcommand \im{i}
\newcommand \dif{{\rm d}}

\begin{document}
\title{Andreev Bound states as a phase sensitive probe of the pairing symmetry of the iron pnictide superconductors.}

\author{Pouyan Ghaemi}
\affiliation{Department of Physics, University of California at Berkeley, Berkeley, CA 94720}

\author{Fa Wang}
\author{Ashvin Vishwanath}
\affiliation{Department of Physics, University of California at Berkeley, Berkeley, CA 94720}
\affiliation{Materials Sciences Division, Lawrence Berkeley National Laboratory, Berkeley, CA 94720}

\date{\today}

\begin{abstract}
A leading contender for the pairing symmetry in the Fe-pnictide high
temperature superconductors is extended s-wave $s_\pm$, a nodeless
state in which the pairing changes sign between Fermi surfaces.
Verifying such a pairing symmetry requires a special phase sensitive
probe that is also momentum selective. We show that the sign
structure of $s_\pm$ pairing can lead to surface Andreev bound
states at the sample edge. In the clean limit they only occur when
the edge is along the nearest neighbor Fe-Fe bond, but not for a
diagonal edge or a surface orthogonal to the c-axis. In contrast to
d-wave Andreev bound states, they are not at zero energy and, in
general, do not produce a zero bias tunneling peak. Consequences for
tunneling measurements are derived, within a simplified two band
model and also for a more realistic five band model.
\end{abstract}

\maketitle

\begin{figure}[htp] \centering
\begin{center}
\vspace{0.1in}\subfigure[]{\label{bz}
\includegraphics[height=4cm]{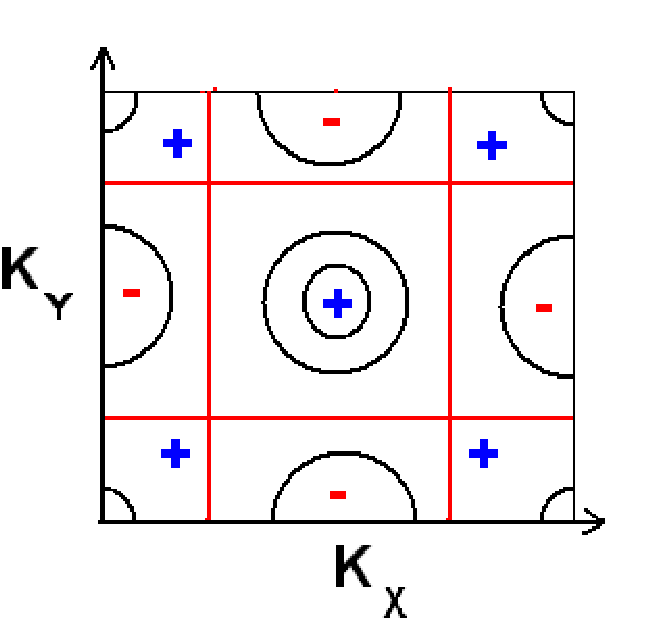} \vspace{0.1in}
} \subfigure[]{
\includegraphics[height=4cm]{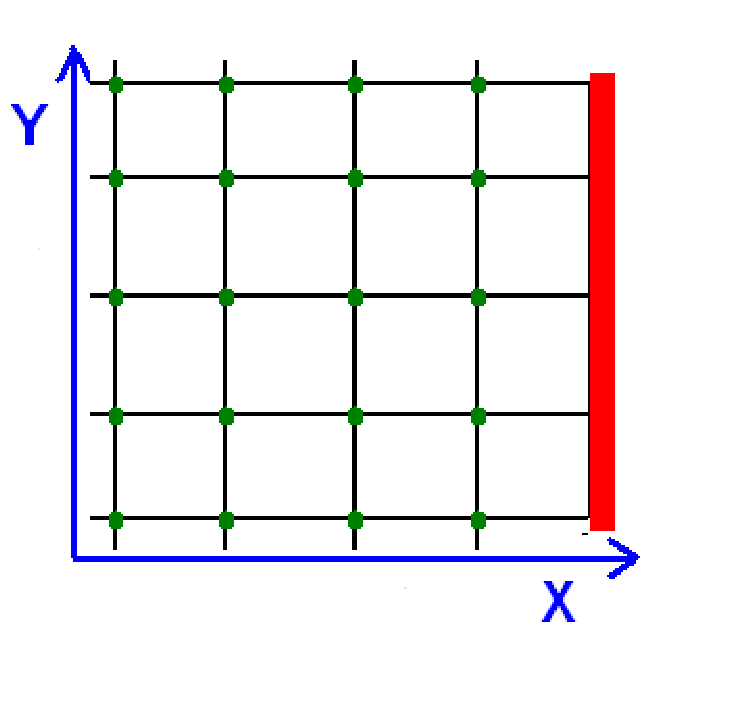}
\vspace{0.1in} \label{pos}}
 \caption{a) Fermi surfaces of Fe pnictides and extended s-wave
pairing sign in the unfolded Brillouin zone b) Real space square
lattice of iron atoms and the [100] boundary (red)}
\label{sc}\end{center}
\end{figure}

Central to the physics of the recently discovered iron pnictide
superconductors, is the nature of the pairing symmetry. Given the
high transition temperatures and proximity to a magnetically ordered
state, a non-phonon mechanism is indicated. Hence, unconventional
pairing symmetry, as often occurs in other such cases like the
cuprates, organic conductors and heavy fermion materials, is to be
expected. Current theories predict a wide variety of possible
pairing states, ranging from p-wave \cite{LeeWen, Zhang} to d-wave
\cite{yao,si}, and also extended s-wave $s_\pm$, a nodeless pairing
state where the pairing changes sign between different Fermi
pockets\cite{Mazin,Chubukov,fa,Wang}. The last is a theoretically
attractive candidate since it is a naturally connected with the
magnetic order observed in the undoped parent materials
\cite{fa,Wang,seo,Chubukov}. While initial experimental results
seemed to favor a pairing state with nodes in the gap function,
recently, the bulk of the experimental evidence has been in favor of
a completely gapped state. For example, ARPES experiments
\cite{arpes} see a full gap around all Fermi pockets, and recent
specific heat measurements are consistent with the absence of nodes
\cite{HHWen}. The extended s-wave state is therefore emerging as an
strong candidate for the pairing symmetry. Hence, it is important to
consider experimental tests that can directly probe such a state. In
contrast to the d-wave symmetry of the cuprates, which changes sign
under rotation and hence can be probed by a class of phase sensitive
experiments \cite{phsens}, the sign change in the $s_\pm$ state is not
accessible by a real space transformation. Instead, as shown in
figure \ref{bz}, the pairing phase changes on moving across the
Brillouin zone. Probing such a sign change is a much more
challenging task. A parametrization of the pairing function that
gives this sign structure on the fermi surfaces is
$\Delta(k_x,k_y)=\Delta_0 \cos k_x \cos k_y$\cite{seo}, where $(k_x,k_y)$
represents the crystal momentum along the Fe-Fe bonds. The pairing
changes sign on crossing the $k_x=\pm \pi/2$ and $k_y=\pm \pi/2$
lines. It is this sign change that we wish to probe.

An interesting consequence of {\em d-wave} pairing in the cuprates
is the presence of Andreev bound states at the edge which leads to a
zero-bias peak in conductivity in superconductor/normal
junctions\cite{kashiwaya,greene,Greene2003162}. The existence of
these modes can be intuitively understood as follows. Consider an
electron with momentum $k$ incident on a surface, where it is
reflected to momentum $k'$. When the pairing function changes sign
in this process $\Delta(k)/\Delta(k')<0$, then the problem is
related to a superconductor-normal-superconductor junction with a
relative phase shift of $\pi$. There it is well known that a midgap
state arises near the chemical potential\cite{Zagoskin}. This leads
to Andreev bound states along those edges of the d-wave
superconductor that reflect particles in the sign reversed way
described above. Note, the Andreev bound states reach all the way
down to the chemical potential (zero energy), and hence are
experimentally detected as a zero bias peak in tunneling
measurements\cite{kashiwaya,greene}. The dependence on edge
orientation can be erased in samples with disordered surfaces.
Nevertheless, in some cases a strong experimental difference has
been observed between 45$^o$ rotated a-b plane
edges\cite{Greene2003162}. The zero bias anomaly does not arise in
c-axis tunneling, as predicted.


In this letter we study the surface Andreev bound states at the edge
of an extended s-wave superconductor. Although our results are more
general, we focus on the models proposed for the pnictide
superconductors. By the intuitive argument described above, we
expect that when an electron is scattered by the edge between Fermi
pockets with opposite sign of pairing, Andreev bound states should
result. We indeed find this to be the case, in the microscopic
models that we study, but with one important distinction from the
d-wave case. The spectrum of Andreev bound states within the bulk
gap does not reach down to zero energy. A zero bias tunneling
anomaly is hence not expected. Nevertheless, the presence of these
Andreev modes will lead to peaks in the tunneling current within the
bulk gap, which should be observable. From figure \ref{bz}, it is
seen that scattering between Fermi surfaces with opposite sign of
pairing occurs for an edge parallel to the nearest neighbor Fe-Fe
bond (along the [100] or [010] directions) but not for the diagonal
edge ([110] direction) or the c-axis edge ([001] direction - we
assume negligible c-axis dispersion). Indeed, we only find bound
Andreev states for the expected edge directions, and these disappear
if the relative  sign of pairing is removed (i.e. with regular
s-wave pairing). The sensitivity to edge orientation is a key
signature of these Andreev surface states. At present, tunneling
experiments on the Fe-pnictide superconductors are restricted to
polycrystalline samples \cite{TesanovicNature}, which makes
comparison to theory difficult. Tunneling experiments on single
crystals can unambiguously identify these states.

We study two microscopic models of the Fe pnictides. The first is a
simplified two band model \cite{raghu,ying}, in which the Andreev
bound states can be studied analytically. The Andreev bound state
energy if found to be controlled by the overlap between the Bloch
wavefunctions of the incident and scattered electronic states, and
satisfies the intuitive requirement that as the wavefunction overlap
decreases, the in-gap state approaches the continuum. With
increasing overlap, the bound state approaches zero energy, which is
achieved if the overlap is perfect. A typical choice of parameters
yields a dispersing in-gap mode whose minimum energy is less than
50\% of the bulk gap. Since the two band model differs in some
subtle but important respects from the LDA determined band
structures \cite{LeeWen}, we have also studied a more realistic
five-band tight binding model \cite{kur}. Here we find qualitatively
similar results. For the particular doping that we study in detail,
the in-gap state is at a higher energy, 92\% of the bulk gap at that
edge momentum. Since a perfectly smooth edge is hard to realize
experimentally, we model edge roughness by allowing for orbital
mixing during scattering (but preserve the edge momentum quantum
number). Within this crude model of a rough edge, the in-gap states
occur at much lower energy (eg. 50\% of the bulk gap value).
Observable signatures in experiments with planar and point contact
tunnel junctions are discussed. In the treatment below, we solve the
Bogoliubov-deGennes equations for superconductor quasiparticles near
the sample edge, with a uniform pairing potential. For weak coupling
superconductors, the pair potential is to be calculated self
consistently, which we do not do here. A self consistent solution is
expected to lead to a smaller pairing potential near the edge, due
to the presence of in-gap states. This is not expected to change our
results qualitatively, and moreover, it is not even clear if such
weak coupling reasoning is justified for these high temperature
superconductors.



{\em Two Band Model:} We first present the the study of the two band
model for pnictides at half filling. The details of this model could
be found elsewhere\cite{raghu}. The band Hamiltonian we consider in
momentum space has the form:


\begin{eqnarray}
H &=& H_0+H_{\Delta} \\
H_0 &=& \sum_{\textbf{k}}
\begin{pmatrix}d^\dagger_{1,\textbf{k},\sigma} & d^\dagger_{2,\textbf{k},\sigma}\end{pmatrix} K(\textbf{k})
\begin{pmatrix} d_{1,\textbf{k},\sigma} \\ d_{2,\textbf{k},\sigma}\end{pmatrix} \label{hamiltonian}\\
H_\Delta &=& \sum_{\textbf{r},\textbf{r'}} \epsilon_{\sigma,\sigma'} \delta_{a,b} \left[ d^{\dagger}_{a,\textbf{r},\sigma}d^{\dagger}_{b,\textbf{r'},\sigma'} \right]+H.c. \label{pairhamilton}
\end{eqnarray}

where $d_1$($d_2$) represents $3d_{xz}$($3d_{yz}$) orbitals of Iron,
$\epsilon$ is the $2 \times 2$ antisymmetric matrix and $\sigma$ is
the spin index. The momentum sum is over the enlarged Brillouin zone
$k_x\in[-\pi,\pi),\ k_y\in[-\pi,\pi)$. The $2\times 2$ matrix
$K(\textbf{k})$ in Ref. \cite{ying}, after rotation into the
$d_{xz,yz}$ basis is:
\begin{align}
&K(k_x,k_y)= 2 t_1(\cos k_x-\cos k_y)\tau_3 + 2(t_2-t_2')\sin k_x \sin k_y\tau_1 \notag\\
& +[ 2(t_2+t_2')\cos k_x\cos k_y+2 t_1'(\cos k_x+\cos
k_y)]\cdot I \label{eq:2-band}
\end{align}
and $\tau_{1,2,3}$ are Pauli matrices and $I$ is the identity
matrix.  Here we present the numerical results for
$t_1=1eV$,$t_2=1.7eV$ and $t'_1=t'_2=0$  at zero doping\cite{ying}.

\begin{figure}[htp]
\includegraphics[width=7.8cm]{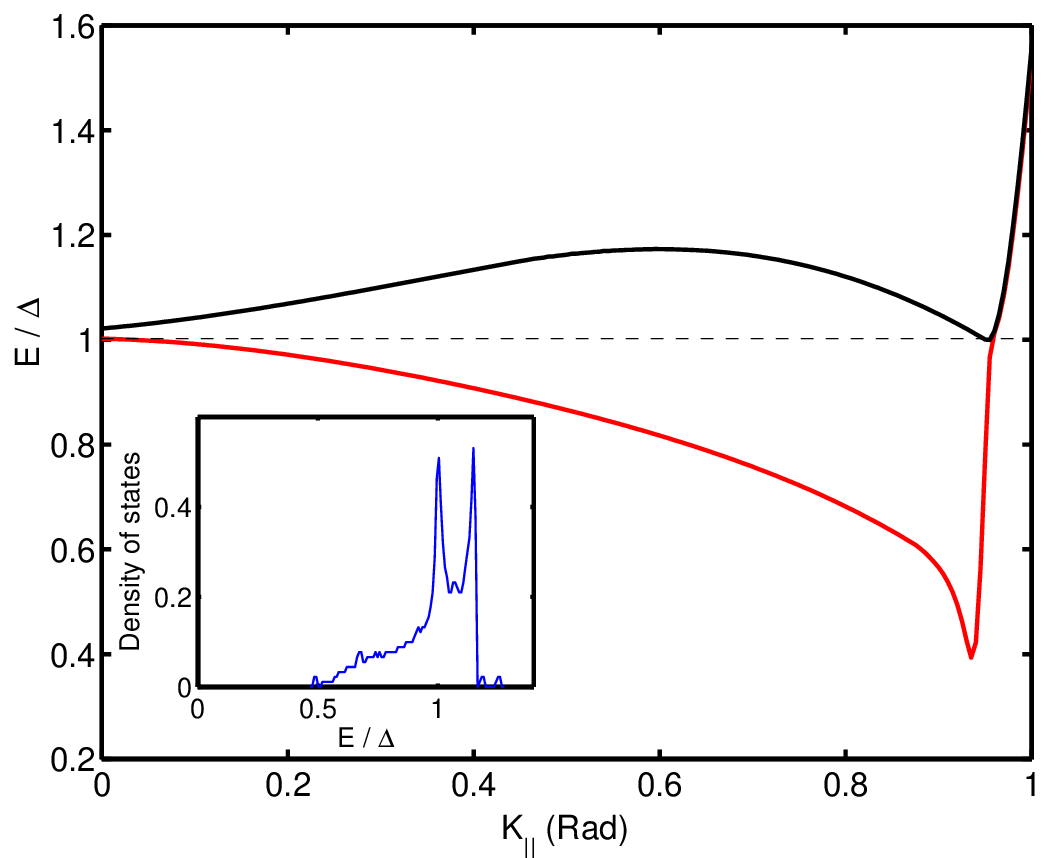}
\caption{Dispersion of Andreev bound states (red) and the bottom of
the continuum (black) as a function of $k_\parallel$, the momentum
parallel to the edge. Energy is scaled by $\Delta$, the minimum
energy gap in the bulk. Inset shows the density of states integrated
over $k_\parallel$.} \label{twoband}
\end{figure}

We numerically diagonalize a 600 site wide lattice, with an edge
along the [100] direction. The momentum $k_\parallel$ along this
direction is conserved and used to label the eigenstates.
Interestingly we see (Figure \ref{twoband}) that a dispersing
mid-gap state is obtained, with energy as low as $39\%$ of the bulk
gap. This is the Andreev surface bound state. For convenience, we
work with a bulk gap of $46\ meV$ which is a few times larger the
experimentally measured gap. We have checked that the energy of the
mid-gap features scale linearly with the gap size over a wide range
hence our results hold when scaled with the bulk gap. The results
were also compared against those obtained from transfer matrix
method\cite{transm} which does not suffer from finite size errors.
The dispersing mode leads to a feature in the density of states
(Figure \ref{twoband} inset), which could potentially be detected in
the tunneling experiments. No surface Andreev states are obtained
with the edge along the [110] direction, or if an onsite s-wave
pairing, with uniform sign in momentum space is used.

In the two band model, one feature of the edge mode as can be seen
in figure \ref{twoband}, is that at transverse momentum $k_{||}=0$
the edge mode merges with the continuum. This arises because the
Bloch wavefunctions on the electron and hole pockets at $k_{||}=0$
are orthogonal (they have opposite eigenvalues under reflection).
Hence, the gap sign changing scattering is forbidden, and no surface
Andreev state appears. On moving away from this special momentum,
the scattering is allowed, leading to Andreev bound state. Our
analytical treatment below captures the relation between
wavefunction overlap and Andreev bound state energy.

\begin{figure}[htp]
\includegraphics[width=4cm]{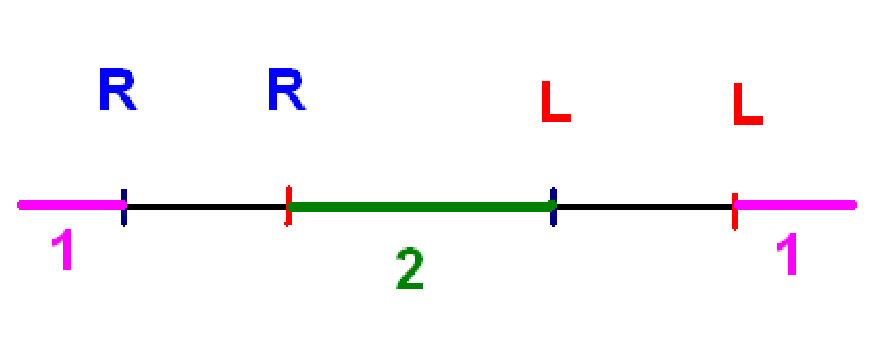}
\caption{Effective one dimensional model of the two band
Hamiltonian, at a fixed $k_\parallel$. Low energy excitations are
labeled as right (R) and left (L) movers. Hole (electron) Fermi
pockets are shown in green (red).} \label{oned}
\end{figure}

We now analytically demonstrate the presence of edge modes as a
result of change of sign of order parameter on the two Fermi
surfaces in the two band model.  For each value of $k_{||}$, the
momentum parallel to the edge, we have a one dimensional
Hamiltonian. The low energy excitations occur close to the Fermi
points, labeled as left mover and right movers (fig. \ref{oned}) for
each Fermi pocket. The low energy electronic field ${\bf
\Psi}_{\sigma}(x)$ written as a two component object to account for
the two orbitals, can be expanded as ${\bf \Psi}_{\sigma}(x)=\sum_n
\psi_{Rn\sigma}(x){\bf u}^R_{n}\ e^{ik_{fn}x}+\psi_{Ln\sigma}(x){\bf
u}^L_n e^{-ik_{fn}x}$, where $n$ labels different Fermi pockets,
$\sigma$ labels physical spin (we suppress $k_{||}$ index from now
on for simplicity) and ${\bf u}^{R/L}_n$ are the two component Bloch
wave functions at the Fermi points.

In these variables, the superconducting Hamiltonian is:


\begin{equation}\label{hamilton}
\begin{split}
H=& \int \dif x
 \sum_{n=1}^{2}  \\
\Big[ & \begin{pmatrix}\psi_{nR\uparrow}^\dagger\ ,&
\psi_{nL\downarrow}^{\vphantom{\dagger}}\end{pmatrix}
\begin{pmatrix}-\im v_n \partial_x & \Delta_{nR}\\
\Delta_{nR}^* & \im v_n \partial_x \end{pmatrix}
\begin{pmatrix}\psi_{nR\uparrow}^{\vphantom{\dagger}} \\ \psi_{nL\downarrow}^\dagger\end{pmatrix} \\
+ &
\begin{pmatrix}\psi_{nL\uparrow}^\dagger ,& \psi_{nR\downarrow}^{\vphantom{\dagger}}\end{pmatrix}
\begin{pmatrix}\im v_n \partial_x & \Delta_{nL}\\
\Delta_{nL}^* & -\im v_n \partial_x \end{pmatrix}
\begin{pmatrix}\psi_{nL\uparrow}^{\vphantom{\dagger}} \\ \psi_{nR\downarrow}^\dagger\end{pmatrix}
\Big]
\end{split}
\end{equation}

Here $n=1,2$ are the hole and electron pockets, while $v_n$ labels
the fermi velocity and $\Delta_n$ is the superconducting gap on the
$n$'th fermi pocket.

We are looking for a localized solution on the edge and consider the
following ansatz to represent the edge mode which diagonalize
Hamiltonian \ref{hamilton}:

\begin{eqnarray}
\nonumber \Phi_{1}&=&\sum_n \int dx \
e^{-\frac{x}{\lambda_n}+ik_{fn}x}
\big(\alpha_{nR\uparrow}\psi_{nR\uparrow}(x){\bf u}^R_{n}
\\ &&\ \ \ \ \ \ \ \ \ \ \ \ \ \ \ \ \ \ \ \
+\alpha^*_{nL\downarrow}\psi^\dagger_{nL\downarrow}(x) {\bf
u}^{L*}_{n} \big)\nonumber \\ \nonumber \Phi_{2}&=&\sum_n \int\ dx\
e^{-\frac{x}{\lambda_n}-ik_{fn}x} \big(\alpha^*_{nR\downarrow}
\psi_{nR\downarrow}^\dagger(x){\bf u}^{R*}_{n} \\ &&\ \ \ \ \ \ \ \
\ \ \ \ \ \ \ \ \ \ \ \ +\alpha_{nL\uparrow}
\psi_{nL\uparrow}(x){\bf u}^L_{n} \big)\nonumber
\end{eqnarray}




They satisfy $[\Phi,H]=E\ \Phi$ with $E<\Delta$ which leads to BdG
equations for $\alpha$'s. At the boundary the particle and hole
amplitude  should separately vanish\cite{Hu}. This leads to the
following boundary equation:
\begin{eqnarray}
\sum_n c_{1n}{\bf u}^{R}_{n}\alpha_{nR\uparrow}+c_{2n}{\bf u}^{L}_{n}\alpha_{nL\uparrow}&=&0 \\
\sum_n c_{2n}{\bf u}^{R*}_{n}\alpha_{nR\downarrow}^*+c_{1n}{\bf
u}^{L*}_{n}\alpha_{nL\downarrow}^*&=&0
\end{eqnarray}

These are set of equations for $c$'s. The condition for existence of
non-zero solution for $c$ determines the bound state energy. The
Bloch function on the two fermi surfaces can then be chosen to have
the simple form: ${\bf u}^R_{1}=(r_1,s_1)^T$, ${\bf
u}^L_{1}=(-r_1,s_1)^T$, ${\bf u}^R_{2}=(r_2,s_2)^T$ and ${\bf
u}^L_{2}=(-r_2,s_2)^T$. We can solve the BdG equations for the bound
state in general, regardless of magnitude of gap on the fermi
surfaces but the expressions are greatly simplified if we assume
that $|\Delta_1|=|\Delta_2|=\Delta$. We present the result for this
case here. When $\Delta_1\Delta_2<0$ we get a pair of bound states
related by time reversal symmetry with energy:

\begin{equation}\label{energy}
E= \pm\Delta \left|\frac{s_2^*r_1 + r_2^*s_1}{s_2r_1-r_2s_1}\right|
\end{equation}

Notice the energy of these states is related to the orbital
structure on two fermi surfaces. With real wavefunctions, these
states occur within the bulk gap if $r_1s_1r_2s_2<0$, which is
satisfied in the two band model of Eq. \ref{eq:2-band}. Only with
complete orbital overlap of ${\bf u}^R_{1}$ and ${\bf u}^L_{2}$ are
they at zero energy. On the other hand, when the gaps have the same
sign (i.e. $\Delta_1=\Delta_2=\Delta$), no localized mode is found
as expected.

\textit{Five band model}: We also preformed a numerical study of a
more realistic five band model\cite{kur} at $18\% $ electron doping.
We again consider a Hamiltonian of the form (\ref{hamiltonian}), but
$d_{\textbf{k},\sigma}$ are now five component spinors. The detailed
form of $K(\textbf{k})$ is given in ref. \onlinecite{kur}.
Superconducting pairing is as in Eqn. \ref{pairhamilton}, but now
with five orbitals. We performed exact diagonalization on lattices
600 sites wide with a [100] edge. While we do not present a general
analytical solution to the five band model edge modes due to its
complexity, we have verified that at $k_\parallel=0$, where the
problem simplifies, the mid-gap state is obtained.

\begin{figure}[htp]
\includegraphics[width=7cm]{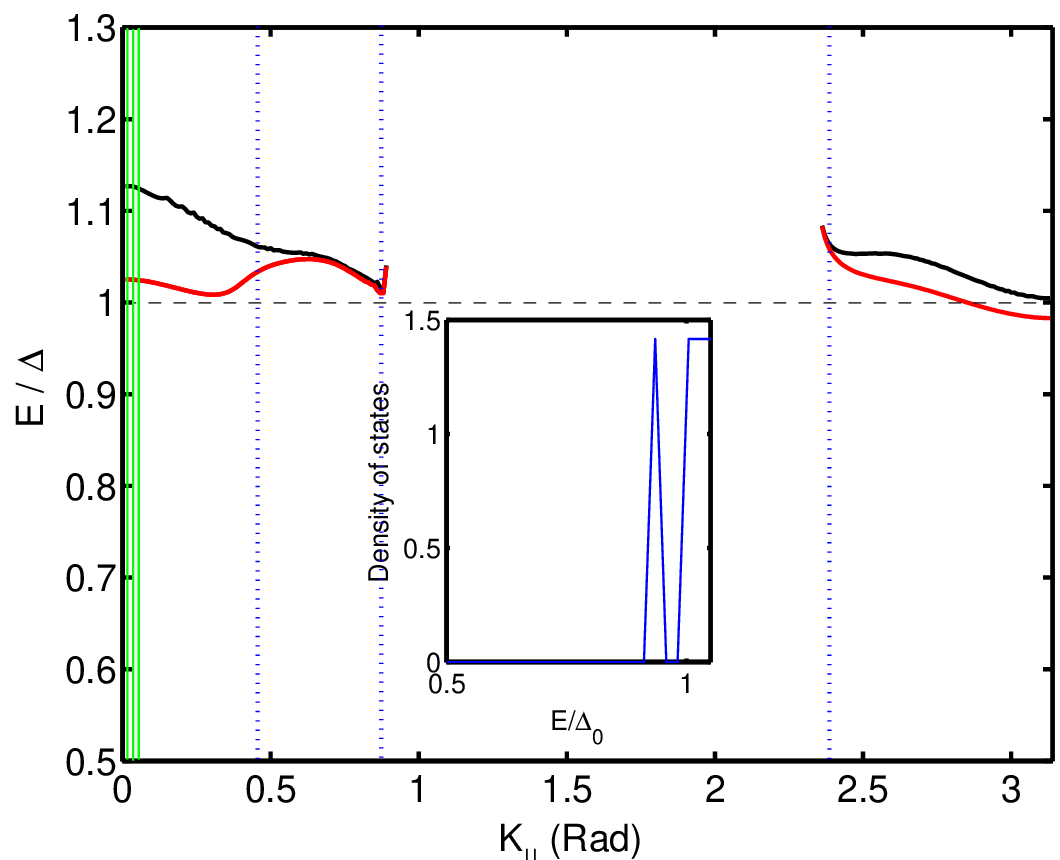}
\caption{Andreev bound state dispersion (red) and bottom of
continuum (black). Inset presents the density of states for
transverse momenta near $k_{||}=0$ (green region). Dotted lines
denote the positions of fermi surfaces} \label{nonpur}
\end{figure}

The dispersion of edge mode as a function of transverse momentum is
shown in Fig.~\ref{nonpur}. Although Andreev bound states are
present, in this case they happen to be close to the continuum.
Around $k_{||}=0$ the Andreev bound state is at $92\%$ of the bottom
of the continuum for that momentum. Note, the precise value of the
edge mode energies depend sensitively on the band structure details,
as well as on doping. Hence, the main value of this result is in
establishing the existence of the Andreev bound states in the five
band model as well. Moreover they also occur for $k_\parallel=0$,
since the wavefunction overlap is nonvanishing at that point, in
contrast to the two band model. In the inset of Fig. \ref{nonpur},
the density of states near $k_\parallel=0$ is shown in units of
$[eV]^{-1}$ per unit-cell. This may in principle be probed by planar
junction tunneling, which probes excitations in a small momentum
cone.

\begin{figure}[htp]
\includegraphics[width=7cm]{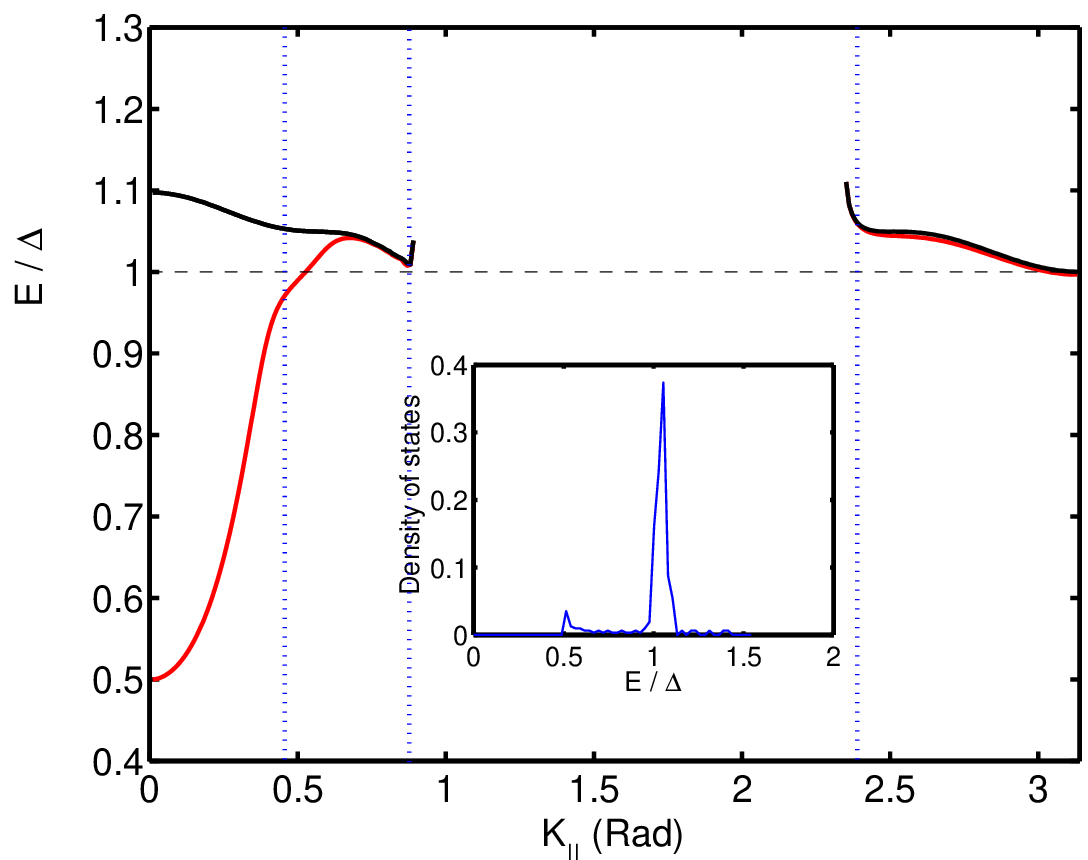}
\caption{Dispersion of in-gap mode (red) and bottom of continuum
(black) after adding a perturbing potential on the edge. Inset
present the density of states. Dotted lines denotes the position of
Fermi surfaces} \label{pur}
\end{figure}

Up to here, we have only considered a clean edge, but this is not
always the case in a realistic experimental setting.
From our analytical treatment it is clear that the energy of the
edge mode depends on the orbital structure on the two fermi surfaces
and so indeed any (non-magnetic) perturbing potential that mixes the
orbitals on the edge can change this mid-gap energy. In
Fig.~\ref{pur} we present one example of the dispersion of mid-gap
states when we have an impurity potential on the edge which
transforms the orbital structure of one fermi surface to the other.
Guided by this expectation we choose the following potential on the
edge.  We choose an impurity potential on the boundary ($x=0$) of
the form $H_{imp}= 0.3\delta_{x,0} [{\bf u}^R_{1} \otimes {\bf
u}^{L}_{2} + h.c.]$ where ${\bf u}^R_{1}$ and ${\bf u}^L_{2}$ denote
the Bloch wave functions for the electrons on the small hole Fermi
surface, and the electron Fermi surface respectively at $k_{||}=0$,
in a gauge where they are purely real. The bound states now reach
down to about $50\%$ of the bulk gap as shown in Figure \ref{pur}.
For technical reasons, we do not consider the most general impurity
potential that would break translation symmetry along the edge and
hence we still retain the momentum $k_\parallel$.

{\em Conclusions:} In this letter we have shown the presence of
Andreev  bound state on the surface of a superconductor with
extended s-wave pairing, which are strongly sensitive to surface
orientation. In contrast to d-wave Andreev bound states, they are
typically at a finite energy. Although the bound state energy is
hard to predict reliably, since it depends in detail on the orbital
structure of Bloch wave functions on the fermi surfaces with
anti-phase paring, the states were found to occur in different
microscopic models of the Fe-pnictides with extended s-wave pairing.
Tunneling experiments on a-b plane edges with planar or point
contacts, or even STM on step edges can potentially detect these
states which, if observed, will be an important step to establishing
the pairing symmetry in these materials. We thank Dung-Hai Lee for
instructive discussions. We acknowledge support from LBNL
DOE-504108.

\bibliography{edgemode10}

\end{document}